# Room-Temperature Ferromagnetism in Fe-doped SnSe Bulk Single Crystalline Semiconductor

Guangqiang Mei, Wei Tan, Xingxia Cui, Cong Wang, Qing Yuan, Yafei Li, Cancan Lou, Xuefeng Hou, Mengmeng Zhao, Yong Liu, Wei Ji, Xiaona Zhang, Min Feng,* and Limin Cao*

[1]*School of Physics and Technology, Key Laboratory of Artificial Micro- and Nano-structures of Ministry of Education, Wuhan University, Wuhan 430072, China*
[2]*Institute of Microstructure and Properties of Advanced Materials, Beijing University of Technology, Beijing 100022, China*
[3]*Beijing Key Laboratory of Optoelectronic Functional Materials & Micro-Nano Devices, Department of Physics, Renmin University of China, Beijing 100872, China*

*Corresponding authors. Email: fengmin@whu.edu.cn; limincao@whu.edu.cn

**ABSTRACT**

The quest for pragmatic room-temperature (RT) magnetic semiconductors (MSs) with a suitable bandgap constitutes one of the contemporary opportunities to be exploited. This may provide a materials platform for to bring new-generation ideal information device technologies into real-world applications where the otherwise conventionally separately utilized charge and spin are simultaneously exploited. Here we present RT ferromagnetism in an Fe-doped SnSe (Fe:SnSe) van der Waals (vdW) single crystalline ferromagnetic semiconductor (FMS) with a semiconducting bandgap of ~1.19 eV (comparable to those of Si and GaAs). The synthesized Fe:SnSe single crystals feature a dilute Fe content of <1.0 at%, a Curie temperature of ~310 K, a layered vdW structure identical to that of pristine SnSe, and the absence of in-gap defect states. The Fe:SnSe vdW diluted magnetic semiconductor (DMS) single crystals are grown using a simple temperature-gradient melt-growth process, in which the magnetic Fe atom doping is realized uniquely using $FeI_2$ as the dopant precursor whose melting point is low with respect to crystal growth, and which in principle possesses industrially unlimited scalability. Our work adds a new member in the family of long-searching RT magnetic semiconductors, and may establish a generalized strategy for large-volume production of related DMSs.

# INTRODUCTION

Many materials platforms for conceptually new information devices and technologies have been and/or are being explored. One such platform is the magnetic semiconductors (MSs) where semiconducting and magnetic properties coexist, and in which the charge and spin degrees of freedom can be manipulated and exploited simultaneously to realize devices of nonvolatile logic-memory functionalities that are not possible in conventional devices.[1-13] Therefore, seeking ferromagnetic semiconductors (FMSs) has long been the focus of fundamental and technological research from both academia and industry, for developing the targeted devices that host intertwined co-spin-charge functionalities.[1-13] An ideal ferromagnetic semiconductor for potentially making such an objective multifunctional device has not yet been found.

From a pragmatic perspective, in order for MSs to function as the building blocks in real and ideal devices in the realm of practical applications, it requires inevitably that the MSs exhibit magnetic critical temperatures ($T_c$, Curie temperature for a ferromagnet) at or above room temperature.[1-13] Besides the utmost important room-temperature (RT) standard of $T_c$, a magnetic semiconductor needs to host a set of essential intrinsic properties of technological relevance, e.g., a suitable semiconducting energy gap ($E_g$), and high stability in air and other working environments, etc. Finally, it is a great advantage and is of particularly practical importance that the MSs are composed of earth-abundant elements with complementary metal oxide semiconductor (CMOS) compatibility, and that a simple and industry-accepted approach is applicable to bulk production of large-size high-quality crystals.

To quest a demandingly pragmatic FMS, numerous materials systems have been continuously explored with sophisticated fabrication processes. Doping transition-metal magnetic elements into the parent nonmagnetic semiconductors to form diluted magnetic semiconductors (DMSs), such as III-V magnetic semiconductors [e.g., (In,Mn)As, (Ga,Mn)As and (Ga,Mn)N], has been extensively pursued using molecular beam epitaxy growth (MBE).[6-13] However, these DMSs normally feature a relatively low Curie temperature $T_c$. A complex DMS $(Ba_{1-x}K_x)(Zn_{1-y}Mn_y)_2As_2$ which is polycrystalline and with $T_c$ up to 180 K was synthesized using the arc-melting solid-state reaction method.[14] A room-temperature magnetic semiconductor of $Co_{28.6}Fe_{12.4}Ta_{4.3}B_{8.7}O_{46}$, which has an amorphous structure and a mobility of 0.1 cm$^2$V$^{-1}$s$^{-1}$, has been implemented by oxidizing the corresponding ferromagnetic Co-Fe-Ta-B metallic glass.[15]

Recently, the magnetic two-dimensional (2D) van der Waals (vdW) materials have emerged into center stage of fundamental and applied materials research. They offer unprecedented opportunities for exploring magnetism and related exotic quantum phases in the 2D limit, for building new-concept devices through manipulating their magnetic and electronic states, and particularly, for easily and rapidly creating various functional heterostructures with nearly perfect surfaces and interfaces, and engineered structures and properties.[1-5,16-22] The exciting discoveries and breakthroughs in this area include the ferromagnetism in atomically thin layers of 2D materials (e.g., $CrI_3$,[23] $Cr_2Ge_2Te_6$,[24] $Fe_3GeTe_2$,[25] $VSe_2$,[26] $MnSe_2$,[27] $Fe:MoS_2$,[28] $Fe:SnS_2$,[29] $V:WS_2$,[30] etc.); the electrostatic tuning of magnetism in atomically thin 2D crystals (e.g., $Fe_3GeTe_2$[25] and $CrI_3$[31-33]); the quantum anomalous Hall effect in quasi-2D Cr- or V-doped $(Bi,Sb)_2Te_3$[34-36] and few-layer intrinsic $MnBi_2Te_4$,[37] etc.; the axion insulator states in magnetic topological insulator $MnBi_2Te_4$[38-40] and magnetic heterostructure of Cr- and V-doped $(Bi,Sb)_2Te_3$ films,[41,42] etc.; a new bulk ferrovalley vdW semiconductor $Cr_{0.32}Ga_{0.68}Te_{2.33}$ with promising spontaneous spin and valley polarization;[43] the prototypical spintronic devices such as a magnetic vdW heterostructured spin filter[44-47] and a spin-orbit torque (SOT) manipulator,[48-52] among an array of others. However, the growth of large-size bulk single crystals of vdW MSs, which may offer an on-demand and clean "LEGO" playground for fundamental studies and device explorations, has been so far scarce.

Herein, we report experimental realization of RT ferromagnetism in Fe-doped SnSe (Fe:SnSe) bulk single crystals which exhibit a $T_c$ of ~310 K. The Fe:SnSe has crystal structure identical to that of pure SnSe, and hosts an $E_g$ of ~1.19 eV, which is comparable to those of the most widely used semiconductors of Si ($E_g$~1.1 eV) and GaAs ($E_g$~1.5 eV). This DMS is implemented via in situ iron-doping from low melting point $FeI_2$ using an industry-standard temperature gradient growth process, which guarantees the low-cost high-volume production of large-size, high-quality single crystals. Given the coveted combination of $T_c$ and an appropriate bandgap, together with the earth-abundant and CMOS-compatible elements in the discovered Fe:SnSe DMS, our work might introduce a new building block for constructing co-spin-charge-functional devices, and engineering hybrid devices into mainstream Si platforms. Furthermore, with the scalable simple growth process in which the low-melting transition-metal halogenides are used as magnetic atom doping precursors, our work may be extended to the production of a variety of MSs with desirable properties, offering the materials foundation for device engineering.

**RESULTS AND DISCUSSION**

SnSe is a simple compound semiconductor consisting of earth-abundant elements with anisotropic layered vdW structure. The schematic unit cell and side and top views of the atomic structures of SnSe are shown in Figure S1, Supporting Information. It hosts remarkable electronic, optoelectronic and thermoelectric properties (which is arguably considered the best thermoelectric crystal known), and a suitable $E_g$ of ~1.0-1.1 eV.[53-56] Inspired by the recent discoveries of ferromagnetism realized by substitutional doping of magnetic elements into the transition-metal dichalcogenides (TMDs) (see, for example, Refs. 28-30), we proposed to dope magnetic elements into the non-magnetic SnSe semiconductor aiming at converting it into a FMS together with other fascinating physical properties inherited from the parent SnSe. For this goal, we adopted the temperature-gradient melt-growth process which is the most cost-effective and simplest approach for bulk crystal growth. It is known that the magnetic transition-metal elements (e.g., Fe, Co, Ni and Mn, etc.) normally have high melting-points, which is drastically unfavorable for melt-growth. To overcome this obstacle, we first converted the high-purity transition-metals (such as Fe in our experiments) into the low melting-point iodides (FeI$_2$ in our experiments) using a simple iodization process. The low melting-point FeI$_2$ (~585 °C) produced was used as the starting dopant precursor in the melt-growth process for the uniform doping of Fe atoms in Fe:SnSe single crystals. This versatile approach in which transition-metal iodides are utilized as the dopant precursors may be applicable to the growth of a variety of DMS crystals with desired semiconducting and magnetic properties.

Figure 1a shows an optical image of the as-synthesized Fe:SnSe single crystals, featuring large sizes (centimeters in three dimensions which are determined by the growth ampule) and smooth layered structure. Figure 1b,c displays X-ray diffraction (XRD) patterns of single crystals and powders (which were made from bulk single crystals by thoroughly grinding in an agate mortar) of the synthesized Fe:SnSe, as well as those of pure SnSe crystals for a comparison. Analyses of XRD data demonstrate that Fe:SnSe possesses an orthorhombic unit cell (lattice constants $a = 11.49 \pm 0.01$ Å, $b = 4.14 \pm 0.01$ Å, $c = 4.43 \pm 0.01$ Å) with *Pnma* space group, which is almost identical to that of pristine SnSe (PDF#48-1224). To clarify the elemental composition in the as-synthesized crystals, we performed X-ray photoelectron spectroscopy (XPS) analyses. A broad Fe $2p_{3/2}$ peak with intensity maximum located at ~707.2 eV can be identified, which corresponds to the binding energy of Fe$^{2+}$ $2p_{3/2}$, verifying the presence of Fe$^{2+}$ state in our Fe:SnSe sample (Figure S2). Combined with the extensive composition analyses using energy-dispersive X-ray spectroscopy (EDX) (see below), we concluded that the as-synthesized Fe:SnSe single crystals had a dilute Fe atomic concentration of <1.0 at% (the atomic ratio of Fe/Sn is lower than 1/50).

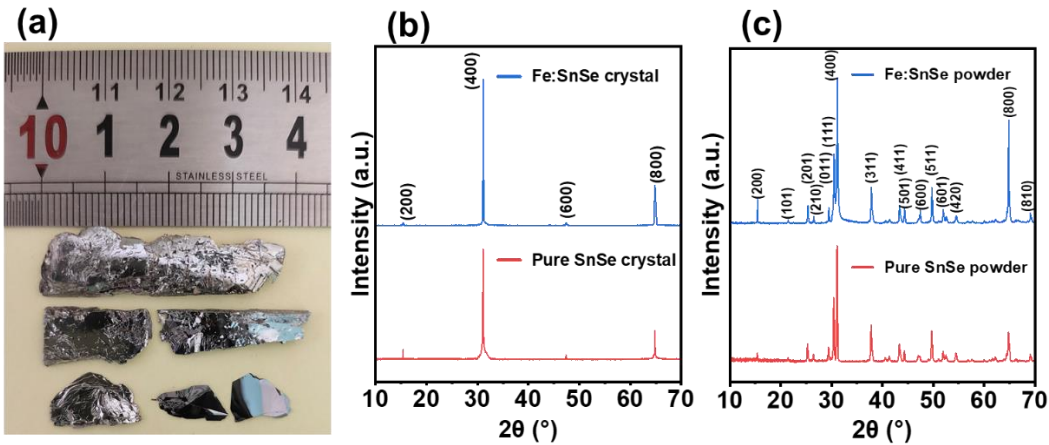

**Figure 1.** Structures of the synthesized Fe:SnSe single crystals. (a) Optical image of the as-synthesized Fe:SnSe single crystals. (b) XRD patterns of single crystals measured along the (100) atomic planes of Fe:SnSe and SnSe. (c) Powder XRD patterns of Fe:SnSe and SnSe crystals. A detailed comparison of the XRD patterns between Fe:SnSe and SnSe verifies that they have identical crystal structure.

With the goal to realize large-size single crystalline DMSs, we characterized the magnetic properties of the synthesized Fe:SnSe and its temperature-dependent characteristic behaviors. In line with our expectations, the obtained Fe:SnSe clearly manifested ferromagnetism at room temperature. Figure 2a shows a set of typical measurements of magnetization versus magnetic field (*M-H*), for which the magnetic field is applied in parallel to the (100) atomic planes (denoted as $H_\parallel$), at different temperatures of 50 K to 350 K. The observation of the well-defined *M-H* hysteresis loops at temperatures below 300 K demonstrates clear evidence for ferromagnetism. The coercive field ($H_c$) at 50 K is ~1960 Oe, while it decreases to ~35 Oe at 300 K (Figure 2b). At 350 K, the magnetization signals show only a diamagnetic background (see inset in Figure 2a). Measuring the remanent magnetization as a function of temperature (Figure 3a) revealed a clear transition that the complete loss of magnetic moment occurred at ~310 K, corroborating a Curie temperature $T_c$ of ~310 K. This is further verified by the measured magnetization as a function of temperature (*M-T*) performed under both field cooling (FC) and zero-field cooling (ZFC) regimes, as shown in Figure 3b, which clearly displays a ferromagnetic to non-magnetic transition at ~310 K.

A series of magnetization measurements performed with the applied magnetic fields perpendicular to the (100) atomic planes (denoted as $H_\perp$) gave similar results (Figure 2c,d). However, the perpendicular ferromagnetic properties of Fe:SnSe exhibit a lower coercivity but a higher saturation magnetization ($M_s$) compared to those measured in plane. For example, $H_c$ of ~820 Oe and $M_s$ of ~1.564 emu/g at 50 K and $H_c$ of ~6.0 Oe and $M_s$ of ~0.095 emu/g at 300 K were measured for $H_\perp$ (Figure 2d). While for $H_\parallel$ (Figure 2b), the measured values were $H_c \approx 1960$ Oe and $M_s \approx 0.533$ emu/g at 50 K and $H_c \approx 35.0$ Oe and $M_s \approx 0.022$ emu/g at 300 K. This indicates the presence of anisotropic ferromagnetic properties in the Fe:SnSe single crystalline ferromagnets (Figure 2). The perpendicular remanence measurements (Figure 3c) demonstrated that the ferromagnetic signal vanished at ~309 K, consistent with the *M-T* measurements under FC and ZFC conditions (Figure 3d). The critical temperature obtained from $H_\perp$ shows good consistency with that measured from $H_\parallel$.

The *M-T* and *M-H* measurements clearly verify that the synthesized crystals remain ferromagnetic above room temperature. Furthermore, the results shown in Figure 2 and Figure 3 demonstrate that our Fe:SnSe crystals possess both in-plane ferromagnetic ordering and inter-plane ferromagnetic coupling with respect to the (100) vdW atomic sheets. Measurements on pure SnSe crystals provided only diamagnetic signals from RT to low temperatures (Figure S3), excluding the assignment of ferromagnetic signal to SnSe. Therefore, we have synthesized large-size Fe:SnSe single crystals in which both the anisotropic in-plane and out-of-plane ferromagnetic ordering persist to above room temperature.

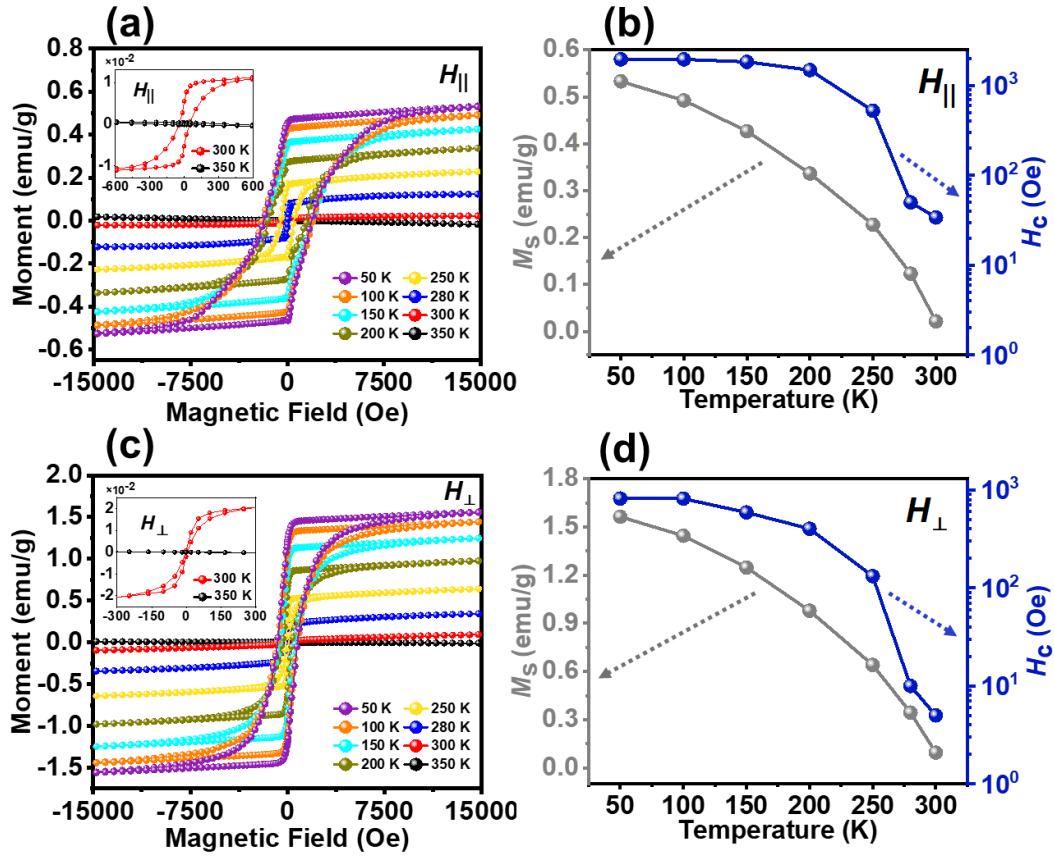

**Figure 2.** Magnetic properties of Fe:SnSe crystals. (a) *M-H* hysteresis loops at different temperatures from 50 K to 350 K. The magnetic field is applied in parallel to the (100) atomic planes of Fe:SnSe. (b) Temperature dependence of the coercive field ($H_c$) and saturation magnetization ($M_s$) measured with $H_{\parallel}$. (c) *M-H* hysteresis loops at different temperatures from 50 K to 350 K measured with the applied magnetic field perpendicular to (100) atomic planes. (d) Temperature dependence of $H_c$ and $M_s$ measured with $H_{\perp}$. The insets in (a) and (c) show the amplified views of the *M-H* hysteresis loops observed at 300 K and 350 K, respectively.

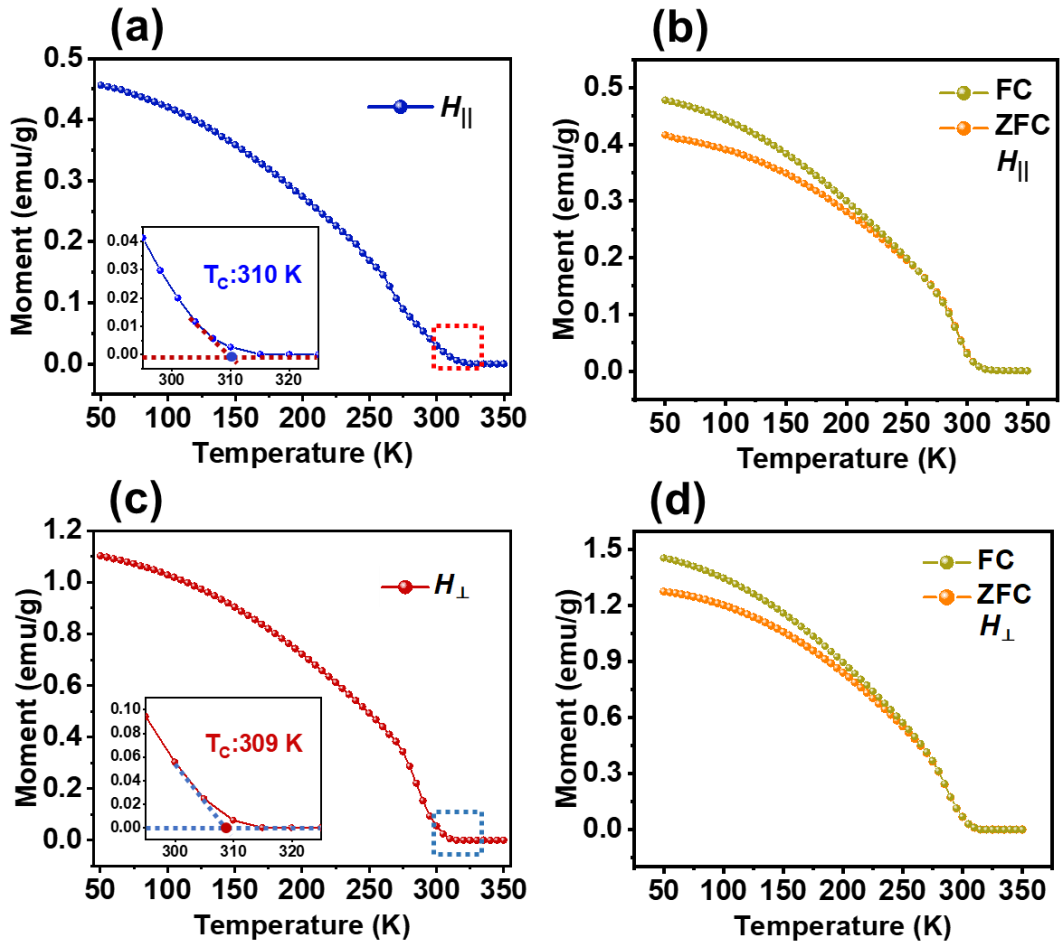

**Figure 3.** Temperature dependent magnetic properties of Fe:SnSe crystals. (a) In-plane remanent magnetization versus temperature. (b) Magnetization as a function of temperature measured under $H_\parallel$ field cooling (FC) and zero field cooling (ZFC) at 1000 Oe. (c) Out-of-plane remanent magnetization versus temperature. (d) Magnetization as a function of temperature under $H_\perp$ field cooling and zero field cooling at 1000 Oe. The remanence measurements were performed by first applying a magnetic field of 20000 Oe at 50 K, and then after a demagnetizing process the remanent magnetic moments were measured with increasing the temperatures from 50 K to 350 K. A close examination of the remanence curves [insets in (a) and (c)] shows clearly a ferromagnetic to non-magnetic transition at $T_c$ of ~310 K. Magnetization measurements under FC and ZFC histories gave consistent results.

We employed high-angle annular dark-field scanning transmission electron microscopy (HAADF-STEM) to characterize the atomic structure of Fe:SnSe crystal. Figure 4a shows an atomic resolution STEM image of our Fe:SnSe. The atomic structure and STEM contrast observed along the [010] zone axis of Fe:SnSe demonstrate an armchair-like atomic configuration which show consistent features with those of pure SnSe.[53-56] The selected area electron diffraction (SAED) patterns shown in Figure 4b and atomic lattice distances revealed in Figure 4a further confirm that our Fe:SnSe possesses crystal structure almost identical to that of SnSe, which are in agreement with an orthogonal structure with lattice parameters of a = 11.37 Å, b = 4.19 Å, and c = 4.44 Å and a space group of *Pnma* (#62). Quantification of the elemental compositions using EDX inside the STEM, as shown in Figure 4c, reveals that the Fe atomic concentration in the Fe:SnSe sample is less than 1.0 at%, which agree with that of XPS analyses.

We acquired atom-resolved elemental mappings of Sn, Se and Fe (shown in Figure 4e-h). The Sn and Se atomic sites agree reasonably well with those revealed in STEM atomic resolution image where larger and brighter spots correspond to Sn atoms (Figure 4d). While, through exhaustive STEM experiments, we found that it were difficult to clearly distinguish the Fe atoms from Sn and Se in the contrast-corrected STEM images and atom-resolved EDX mappings though Fe ($Z = 26$) has a smaller atomic number than Sn ($Z = 50$) and Se ($Z = 34$) atoms. This may be because the Fe content in our Fe:SnSe is too low to be clearly and directly identified. However, EDX mapping elemental distributions at low magnifications in the STEM demonstrate that Fe atoms are clearly visible over large area and distribute rather uniformly in the crystals (Figures S4 and S5). Extensive STEM imaging and EDX mapping studies confirm that there is no Fe atom clustering and/or Fe-rich phase segregation taking place in either the vdW gaps or lattice sites of the Fe:SnSe crystal.

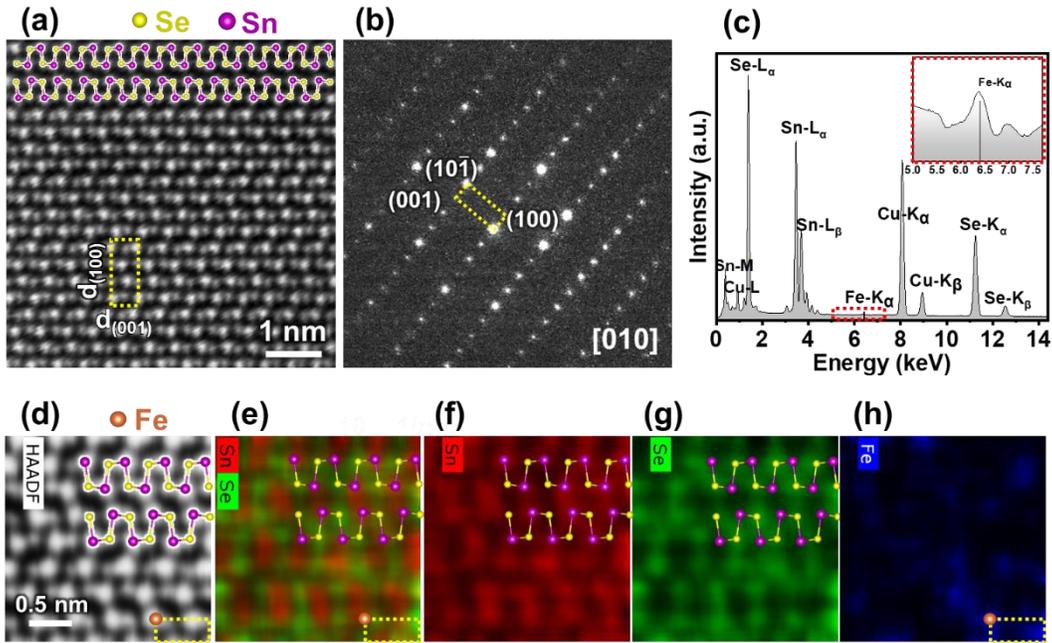

**Figure 4.** STEM characterizations of Fe:SnSe crystal. (a) High resolution STEM image of Fe:SnSe acquired along the [010] zone axis. The ball-and-stick model of Sn-Se atomic armchair rows is superposed on the STEM image to locate Sn and Se atoms. The measured lattice atomic plane spacings are $d_{(100)} = 11.47$ Å and $d_{(001)} = 4.50$ Å. (b) SAED pattern corresponding to (a). The measured lattice parameters for (100), (001) and (101) atomic planes are $d_{(100)} = 11.48$ Å, $d_{(001)} = 4.44$ Å, and $d_{(101)} = 4.17$ Å, respectively. (c) EDX spectra of Fe:SnSe. The inset shows the zoomed-in EDX spectra around the Fe $K_\alpha$ peak. (d-h) Atomic resolution STEM image and the corresponding elemental mappings of Sn, Se, and Fe, respectively. The brown balls and the yellow-dotted rectangles in (d,e,h) are marked to try to locate the position of one of the Fe atoms which shows a relatively better contrast in the atomic elemental mapping of Fe in (h). The ball-and-stick models of Sn-Se atomic armchair rows are superposed on the STEM and the corresponding elemental mapping images.

In our control experiments, we have prepared Fe:SnSe single crystals with different atomic ratios of the starting precursor $FeI_2$ from 1.0 at% to 5.0 at%. However, it was found that atomic doping concentration of Fe had a limited tunability, which almost did not exceed 1.0 at% in all the products. This may be attributed to the high energy barrier for Fe to replace other metal atoms in 2D vdW materials.[28,57,58] Considering that Fe is a transition metal element, we suggest that Fe might substitute the Sn host site, similar to those demonstrated in other Fe-doped transition metal dichalcogenides (TMDs) vdW materials.[28,29,57,58] Indeed, our preliminary density functional theory (DFT) calculations showed that that a substitution of Sn host site with Fe was thermodynamically favored (the theoretical results will be described elsewhere). This is further supported by our atomic-scale measurements of local atomic and electronic properties of Fe:SnSe surfaces using scanning tunneling microscopy/spectroscopy (STM/STS) and qPlus atomic force microscopy (qPlus AFM).

Figure 5a shows a typical large-area STM topographic image of freshly cleaved Fe:SnSe(100) surfaces. The most obvious feature is that there exist much more defects in Fe:SnSe, which present distinct STM contrasts, compared with pure SnSe (Figure S6a). According to the STM contrasts at different biases, the defects are generally categorized into defect 1 to 4 as marked in Figure 5a. Among these defects, defect 2 is dominant. Closely examining the atomic structure of defect 2 revealed that it might be originated from Fe substitution of Sn. Though STM topographic image appears that there is a lattice distortion around defect 2 (Figure S7), the atom-resolved qPlus AFM image (Figure 5b) reveals that atomic plane is smoothly continuous around defect 2 and the Sn site is replaced by a smaller atom (the one at the center of the white-dotted circle in Figure 5b). This type of defects is the uniquely characteristic one in Fe:SnSe; no such defect is observed in pure SnSe. This supports our deduction that it originates from the substitutional Fe at the Sn site.

The lattice constants measured by both STM and AFM on Fe:SnSe(100) surface are consistent with those of pure SnSe, which is in agreement with our XRD and TEM results. Importantly, we did not observe clusters through thorough STM examinations on the Fe:SnSe(100) crystal surfaces freshly cleaved under ultrahigh vacuum in the STM preparation chamber; this is consistent with STEM imaging and EDX mapping results. The experimental data suggests that the formation of clusters seems unlikely to happen in the vdW inter-layer space of SnSe atomic planes.

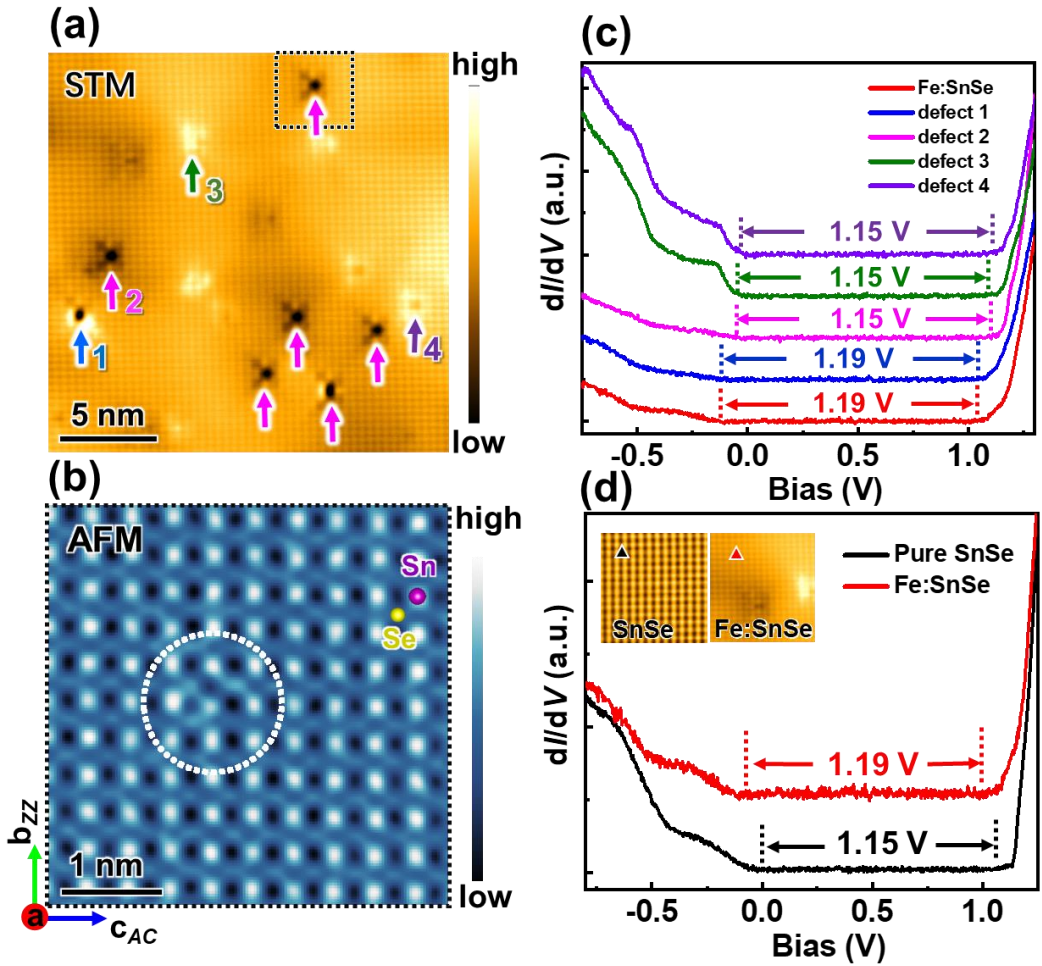

**Figure 5.** STM and AFM characterizations of the Fe:SnSe crystals. (a) Large-area STM image revealing the high density of defects, which are generally categorized into defect 1 to 4 marked by the arrows with different colors. STM imaging parameters: bias voltage $V_b = -1.3$ V, tunneling current $I_t = 10$ pA. (b) A typical atom-resolved qPlus AFM image ($\Delta f = -5.95$ Hz) acquired from the region marked by the black-dotted rectangle enclosing the characteristic type 2 defect in (a). The surface lattice constants are measured to be $4.48 \pm 0.01$ Å and $4.15 \pm 0.01$ Å along the armchair (*AC*) and zigzag (*ZZ*) directions, respectively. The lattice constants are consistent with those measured on pure SnSe with $c_{AC} = 4.44 \pm 0.01$ Å and $b_{ZZ} = 4.16 \pm 0.01$ Å (Figure S6 and S7). The white-dotted circles in (b) and in Figure S7 enclose the same defect which represents a characteristic defect 2 in (a). (c) STS spectra measured from smooth Fe:SnSe surface (the red curve) where is free of defects, and from the sites of defects 1-4 with corresponding colors marked in (a), showing its p-type semiconductor characteristic with an $E_g$ of ~1.19 eV. (d) A comparison of STS spectra between Fe:SnSe and pure SnSe. The measured p-type $E_g$ for pure SnSe is ~1.15 eV.

The black and red triangles in the inset STM images mark the regions on SnSe(100) and Fe:SnSe(100) surfaces from where the characteristic STS spectra (black curve for SnSe and red curve for Fe:SnSe) have been acquired.

Scanning tunneling spectroscopy (STS) spectra (Figure 5c,d) of the Fe:SnSe highlight its p-type semiconductor characteristics with measured $E_g$ of ~1.19 eV, which is a little larger than that of pure SnSe ($E_g$ ~1.15 eV in our STM measurements). This energy bandgap is comparable to those of the most widely used industrial semiconductors of Si ($E_g$ ~ 1.1 eV) and GaAs ($E_g$ ~ 1.5 eV). Of particular importance here is that, as shown in Figure 5c, the STS spectra acquired from Fe:SnSe manifest no presence of in-gap defect states no matter where they are measured (either acquired from the smooth areas or just from the different defect sites). This characteristic also verifies that the substitutional doping of Fe atoms takes place, since either interstitial and/or intercalated Fe atoms would probably introduce atomic in-gap defect states within SnSe's pristine bandgap. The absence of in-gap defect states and the fact of bandgap values keeping constant throughout the crystal feature a unique character of the synthesized Fe:SnSe crystal, which may offer a compelling advantage for its applications in electronic and optoelectronic devices.

We note that further theoretical and experimental work is needed to understand the origin of RT ferromagnetism as well as to study the atomic and electronic structures in the synthesized Fe:SnSe single crystals. However, we expect that, together with the RT ferromagnetism, an appropriate $E_g$ of ~1.19 eV, the absence of in-gap defect states, and other fascinating properties possibly inherited from the parent SnSe semiconductor (such as the record-setting thermoelectric performance), the obtained vdW Fe:SnSe may provide a materials foundation for future explorations. For example, considering its wafer-volume production and atomically smooth vdW surface and interface, it can be used directly as a magnetic semiconductor substrate for the fabrication of various exotic electronic and optoelectronic devices, which offer an integrated solid-state magnetic environment for achieving and manipulating unique functionalities. When exfoliated into atomically thin layers, it can serve as the magnetic tunneling barriers and/or as an essential component to construct various vdW heterostructures for building high-performance new-concept spintronic devices and/or artificial quantum materials. These coveted opportunities to be explored might make the synthesized Fe:SnSe RT ferromagnetic semiconductor attractive as the versatile "LEGO" components for future materials and device engineering.

**CONCLUSIONS**

In summary, using a simple temperature-gradient melt-growth process, we have grown large-size (centimeters in three dimensions) vdW Fe:SnSe bulk single crystals in which the intrinsic semiconducting ($E_g$ ~1.19 eV) and RT ferromagnetic ($T_c$ ~310 K) properties coexist. In the growth process, we use the low melting point $FeI_2$ as the dopant precursor to achieve an effective Fe atomic doping into the pristine SnSe semiconductor, which is free from expensive facility and complicated techniques. This scalable process may be applicable to the growth of a variety of MSs and other exotic quantum materials that need doping with high melting-point transition metal atoms, which is otherwise difficult to achieve if pure transition-metal elements are used. We have grown Fe-doped $SnSe_2$ (Fe:$SnSe_2$) DMS single crystals with centimeter-scale sizes, which host a $T_c$ of ~320 K, higher than that of Fe:SnSe, using this cost-effective and scalable growth process.

**METHODS**

**Synthesis of Fe:SnSe and SnSe Single Crystals.** The iron-doped SnSe (Fe:SnSe) and pristine SnSe single crystals were grown using a modified temperature-gradient melt-growth method. In the growth of single crystals, high purity (99.9999%) Sn and Se granules were used as the starting materials. For Fe:SnSe single crystals, besides Sn and Se, high purity $FeI_2$ (purity better than 99.95%) was used as the Fe doping precursor. The $FeI_2$ was synthesized via a simple iodization reaction of iron powders (purity 99.99%) and iodine granules (purity 99.99%) at a temperature of 520 °C, which were vacuum-sealed in a quartz tube at $5\times10^{-5}$ Torr. Sn, Se and FeI$_2$ granules with the designed stoichiometric atomic ratios and a total weight of about 30 g were loaded into a quartz ampoule with an inner diameter of 11 mm. Then the ampoule was evacuated to $<5\times10^{-5}$ Torr and sealed. The primary ampoule was inserted in another quartz tube with inner diameter of 16 mm, and was evacuated and sealed to protect the sample and ampoule. The double-sealed quartz tubes were loaded into a tubular furnace at a 15° angle from the horizontal plane. The samples in the furnace was slowly heated to 980 °C over 30 hours, soaking at this temperature for 48 hours, and then cooling from 980 to 500 °C with a precisely controlled cooling rate of 1 °C h$^{-1}$. After cooling the furnace to room temperature, the synthesized SnSe and Fe:SnSe single crystals were taken out from the quartz ampoule and used for the experiments.

**XRD, STEM and XPS Characterizations.** X-ray diffraction analyses were carried out using a Rigaku SmartLab SE X-ray diffractometer with Cu K$\alpha$ radiation. HAAD-STEM observations were carried out using a Talos F200X G2 probe-corrected scanning transmission electron microscopy equipped with an energy-dispersive X-ray spectroscopy (EDX). The STEM analyses were performed at an acceleration voltage of 200 kV with a spatial resolution of ~1.4 Å. XPS measurements of both of SnSe and Fe:SnSe single crystal samples were carried out using an ESCALAB250Xi XPS spectrometer, where Al K$\alpha$ X-rays were used.

**STM/STS and qPlus AFM Measurements.** STM, qPlus AFM, and spectroscopy experiments were carried out in an ultrahigh vacuum (UHV) low temperature STM system (CreaTec). STM topographic images were acquired in constant-current mode. The d$I$/d$V$ spectra were measured using the standard lock-in technique with a bias modulation of 8.0 mV at 711.333 Hz. The STM tips were chemically etched tungsten, which were further calibrated spectroscopically against the Shockley surface states of cleaned Cu(111) or Au(111) surfaces before performing measurements on the freshly cleaved SnSe and Fe:SnSe single crystals (the Fe:SnSe and SnSe crystals were cleaved in-situ in a preparation chamber under ultrahigh vacuum at room temperature). For the qPlus AFM measurement, the tip was decorated with CO by picking up a single molecule from an Ag(100) surface. The parameters for picking up CO were sample bias of $V_b = 40$ mV with tunneling current $I_t = 100$ pA. The AFM imaging was performed by frequency modulation (FM-AFM) with a constant amplitude of $A = 100$ pm. The resonance frequency of the AFM probe was $f_0 = 24.5$ kHz and its quality factor Q was 47962 at ~5.0 K.

**Magnetization Measurements.** Temperature- and magnetic field-dependent magnetization measurements were performed using the vibrating sample magnetometers (VSM) in a superconducting quantum interference device (MPMS3, Quantum Design) and a physical property measurement system (PPMS, Quantum Design). The measurements were carried out in two types of magnetic fields applied with respect to the orientation of crystals. One was that the magnetic field was applied in parallel to the (100) atomic planes (denoted as $H_\parallel$). Another was applied perpendicular to the (100) atomic planes (denoted as $H_\perp$). The Fe:SnSe single crystal slices with weight of ~5 mg were cleaved from the as-grown crystal ingot for the magnetic measurements. The freshly cleaved slice was glued on a standard quartz sample holder using wax for magnetization measurements under parallel $H_\parallel$ field, while a special quartz holder on which the crystal slices were placed horizontally was used for the measurements under perpendicular $H_\perp$.

## ACKNOWLEDGMENTS
We gratefully acknowledge the productive discussions with J. Yuan, M. He, F. X. Zhang, R. Yu, L. Mao. This work is supported by the Strategic Priority Research Program of Chinese Academy of Sciences (Grant No. XD30000000), the National Key R&D Program of China (Grant Nos. 2018YFE0202700, 2017YFA0303500), the National Natural Science Foundation of China (Grant No. 11774267).

## REFERENCES
(1) Yang, H.; Valenzuela, S.O.; Chshiev, M.; Couet, S.; Dieny, B.; Dlubak, B.; Fert, A.;